\documentclass[12pt, onecolumn]{IEEEtran}

\usepackage{graphicx}
\usepackage{psfrag}
\usepackage{amsfonts}
\usepackage{algorithm}
\usepackage{algorithmic}
\usepackage{subfigure}
\usepackage{textcomp}
\usepackage{setspace}

\def\ve#1{{\mathchoice{\mbox{\boldmath$\displaystyle #1$}}%
              {\mbox{\boldmath$\textstyle #1$}}%
              {\mbox{\boldmath$\scriptstyle #1$}}%
              {\mbox{\boldmath$\scriptscriptstyle #1$}}}}

\newcommand{\BER}{\mathrm{BER}}
\newcommand{\FER}{\mathrm{FER}}

\newcommand{\efkt}{\mathrm{e}}
\newcommand{\expect}{\mathrm{E}}
\newcommand{\mentrop}{{\mathrm{e}}_M}
\newcommand{\binentrop}{{\mathrm{e}}_2}

\newcommand{\ld}{\log_2}

\begin{document}

\title{{\LARGE{The Lowest-Possible BER and FER for any Discrete Memoryless Channel with Given Capacity}}}

\author{
\authorblockN{Johannes B. Huber and Thorsten Hehn\\
\authorblockA{Institute for Information Transmission\\ University of Erlangen-Nuremberg, Germany}
}}
\maketitle
\doublespacing
\begin{abstract}
We investigate properties of a channel coding scheme leading to the minimum-possible frame error ratio when transmitting over a memoryless channel with rate $R>C$. The results are compared to the well-known properties of a channel coding scheme leading to minimum bit error ratio. It is concluded that these two optimization requests are contradicting. A valuable application of the derived results is presented.
\end{abstract}

\section{Introduction}
\label{sec:introduction}

We consider coded data transmission over a memoryless channel with given capacity $C$ for the case that the rate of the channel exceeds the channel capacity. This is a typical situation for a component code in a concatenated coding scheme \cite{brink01}. The properties of a channel coding scheme with minimum average bit error ratio ($\BER$) have been discussed in several papers (e.g.\ \cite{huettingeretal02} and references therein). In this paper, we focus on schemes with minimum average frame error ratio ($\FER$). The work presented in this paper is threefold. As a first contribution, we discuss a coding scheme optimal w.r.t.\ $\FER$ and use rate-distortion theory to derive the properties of the end-to-end channel. Second, we present a possible application of these findings. This is a lower bound on the rate when both the channel capacity and the tolerated average frame error ratio are specified. This leads to the most important and third contribution, the insight that minimum $\BER$ and minimum $\FER$ are contradicting targets which cannot be obtained by a single channel coding scheme. We show the consequences for the $\BER$ when the channel coding scheme is optimized w.r.t.\ to the $\FER$ and vice versa. The paper is organized as follows. Section \ref{sec:transmission_setup} provides necessary definitions and describes the transmission system. In Section~\ref{sec:converse} we repeat the converse to the channel coding theorem which identifies a lower bound for reliable transmission. Section \ref{sec:lowest_ber} briefly repeats the properties of channel coding scheme with minimum $\BER$, which was derived in \cite{huettingeretal02}. In Section \ref{sec:lowest_fer} ideal channel coding w.r.t.\ $\FER$ is introduced. An application as well as performance results for the non-optimized error ratio are shown in Section~\ref{sec:possible_application}.

\section{Transmission Setup and Definitions}
\label{sec:transmission_setup}

\begin{figure}[h!]
\begin{center}
\psfrag{Bin}[c][c][0.75]{{\footnotesize{Binary}}}
\psfrag{So}[c][c][0.75]{{\footnotesize{Source}}}
\psfrag{Enc}[c][c][0.75]{{\footnotesize{Encoder}}}
\psfrag{RR}[c][c][0.75]{{\footnotesize{$R=\frac{k}{n}$}}}
\psfrag{Ch}[c][c][0.75]{{\footnotesize{DMC}}}
\psfrag{Cap}[c][c][0.75]{{\footnotesize{Capacity $C$}}}
\psfrag{Dec}[c][c][0.75]{{\footnotesize{Decoder}}}
\psfrag{Si}[c][c][0.75]{{\footnotesize{Sink}}}
\psfrag{Etec}[c][c][0.75]{{\footnotesize{End-to-end channel}}}
\psfrag{U}[c][c][0.75]{{\footnotesize{$\ve{u}$}}}
\psfrag{X}[c][c][0.75]{{\footnotesize{$\ve{x}$}}}
\psfrag{Y}[c][c][0.75]{{\footnotesize{$\ve{y}$}}}
\psfrag{V}[c][c][0.75]{{\footnotesize{$\ve{v}$}}}
\psfrag{Ud}[c][c][0.75]{{\footnotesize{$\hat{\ve{u}}$}}}
\includegraphics[scale=0.6]{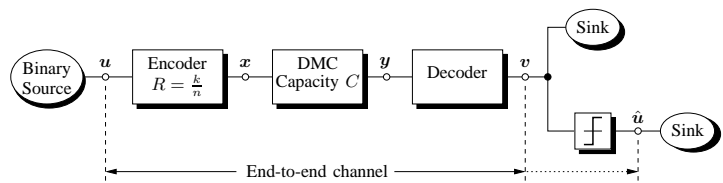}
\caption{\label{fig:transmission_scenario}Transmission scenario for signaling over a DMC of capacity $C$}
\end{center}
\end{figure}

An information source delivers source symbols $u[\ell]$, $\ell=1,\dots,k$, from a binary alphabet. These symbols are realizations of the random variables $U[\ell]$ which are assumed to be independent of each other and have identical distributions. In other words, $H(U[\ell])=1$ holds, where $H(\cdot)$ denotes the entropy of a random variable. In the following, we denote the $j$-th element of the set of possible source words by $\ve{u}^{(j)}$, $j=1,\dots,2^k$ and $u[\ell]$, $\ell=1,\dots, k$ denotes the $\ell$-th entry in a vector $\ve{u}$ of length $k$. An encoder of rate $R=k/n$ is used to transform binary source vectors $\ve{u}$ of length $k$ into vectors $\ve{x}$. These vectors contain channel symbols and are of length $n$.
The vectors $\ve{x}$ are transmitted over the channel and received as vectors $\ve{y}$ of length $n$, cf.\ Figure~\ref{fig:transmission_scenario}. Note that we do not assume any special properties of the channel except for being discrete and memoryless (discrete memoryless channel, DMC) and meeting the capacity $C$. A corresponding channel decoder uses the received vector $\ve{y}$ to generate soft-output estimates of $\ve{u}$, denoted by $\ve{v}$. Binary quantization of $\ve{v}$ yields $\hat{\ve{u}}$. As stated above, we are interested in channel coding schemes that minimize the $\BER$ and $\FER$, respectively, when measured over the end-to-end channel. Here, the end-to-end channel corresponds to the channel transmitting $\ve{u}$ to $\ve{v}$ if soft-decision output is required and $\hat{\ve{u}}$ otherwise, cf.\ Figure~\ref{fig:transmission_scenario}.

We define the average bit error ratio for a given position $\ell$ as $\mathrm{BER}_{\ell} =\Pr(\hat{U}[\ell] \neq U[\ell])$, $\ell=1,\dots, k$ and the average bit error ratio in a codeword as 
$\BER=\frac{1}{k}\sum\limits_{\ell=1}^{k} \mathrm{BER}_{\ell} = {\expect_{\ell}}\{\mathrm{BER}_{\ell}\}$.
Additionally, a {\emph{tolerated}} average bit error ratio $\BER_{\mathrm{T}}$ is introduced. This ratio is technically equal to $\BER$, as it is also measured between $\ve{u}$ and $\hat{\ve{u}}$, but $\BER_{\mathrm{T}}$ is a user-defined threshold variable. Henceforth, we tacitly assume $\BER_{\mathrm{T}}\leq 0.5$.
Similarly, we define the average frame error ratio as the probability that the received frame differs from the transmitted one, i.e.\
$\FER=\Pr(\ve{U}\not=\hat{\ve{U}})$
where equality of two vectors is given if all elements in the two vectors are equal.
Alike $\BER_{\mathrm{T}}$, we consider the tolerated average frame error ratio $\FER_{\mathrm{T}}$.

$I(\ve{U};\ve{V})$ denotes the total mutual information between vectors (frames) $\ve{u}$ and $\ve{v}$ while $I(U[\ell];V[\ell])$ is the mutual information between an individual pair of input and output symbols. Additionally, we define the average mutual information transmitted in a frame of $k$ symbols,
$\bar{I}(U;V)=\frac{1}{k}\sum_{\ell=1}^kI(U[\ell];V[\ell])$.

\section{Converse to the channel coding theorem}
\label{sec:converse}

We repeat the converse to the channel coding theorem as stated in \cite[Ch.\ 4]{gallager68}. This theorem marks the starting point for both our considerations on the lowest $\BER$ and $\FER$. In \cite[Ch.\ 4]{gallager68} the transmission of a sequence of source digits is discussed. Note that these digits can represent both single information symbols as well as complete source words. We represent this distinction by different alphabets and as a consequence we denote the length of a sequence of source symbols by $L$. As each digit can be taken from an arbitrary alphabet, we denote the sequence of source digits by a sequence of vectors, $\ve{u}_1^L=[\ve{u}_1,\dots, \ve{u}_L]$. The sequence of channel digits and received digits, both of length $N$, are denoted by $\ve{x}_1^N=[\ve{x}_1,\dots, \ve{x}_N]$ and $\ve{y}_1^N=[\ve{y}_1,\dots, \ve{y}_N]$, respectively. Please note that the cardinalities of the sets of channel input and output variables do not have to be specified, the mutual information $I(\ve{X}_1^N;\ve{Y}_1^N)$ is sufficient. In the following, we consider the general error event that the source digit and the estimated digit do not coincide and denote the probability by $P_e$. Identifying the source digits from alphabets of size $M=2$ and $M=2^k$ allows us to deduct information on the average error ratios of interest, i.e. the $\BER$ and the $\FER$, respectively.

We start by Equation (4.3.20) from \cite{gallager68} (Fano's inequality), which is a lower bound on the error probability $P_e$ and reads

\begin{equation}
\mentrop(P_e)\geq \frac{1}{L}H(\ve{U}_1^L)-\frac{1}{L}I(\ve{X}_1^N;\ve{Y}_1^N).
\label{eq:common_pe}
\end{equation}
Here, $\mentrop(\cdot)$ is the $M$-ary entropy function, 
$\mentrop(p)=\binentrop(p)+p\log_2(M-1)$, and $\binentrop(\cdot)$ is the usual binary entropy function $\binentrop(p)=-p\log_2(p)-(1-p)\log_2(1-p)$. Further, $M$ denotes the size of the symbol alphabet. 

Let us first assume that $\ve{u}_1^L$ is a vector of binary symbols of length $L$, i.e.\ $L=k$ and $N=n$. In this case, $P_e$ coincides with the $\BER$ and $\frac{1}{k}H(\ve{U}_1^k)=1$ holds. Together with $I(\ve{X}_1^n;\ve{Y}_1^n)\leq nC$ due to the memoryless channel and the capacity maximum of mutual information, the well known lower bound on $\BER$ results:

\begin{equation}
{\mathrm{e}}_2(\BER)\geq 1-\frac{n}{k}C = 1-\frac{C}{R}.
\label{eq:bound_gallager_ber}
\end{equation}

%

%

%

As mentioned above, we regard the source sequence $\ve{u}$ for a lower bound on $\FER$ as one symbol out of a $2^k$-ary alphabet, and the variables $\ve{x}$ and $\ve{y}$ represent an entire codeword and received word, respectively. In block coding, each source sequence is encoded into one codeword and subsequent codewords are mutually independent. Therefore, $L=N$ holds and Equation~(\ref{eq:common_pe}) reads
$$ 
{\mathrm{e}}_{2^k}(\FER)\geq \frac{1}{L}H(\ve{U}_1^L)-\frac{1}{L}I(\ve{X}_1^L;\ve{Y}_1^L) = H(\ve{U}) - I(\ve{X};\ve{Y}).
$$
With $H(\ve{U})=k$ and $I(\ve{X};\ve{Y})\leq nC$, we obtain the corresponding result to Equation~(\ref{eq:bound_gallager_ber}) for $\FER$,

\begin{equation}
{\mathrm{e}}_{2^k}(\FER)\geq k-nC = k(1-\frac{C}{R}).
\label{eq:bound_gallager_fer}
\end{equation}

%
%
%
It is worth mentioning that Equation~(\ref{eq:bound_gallager_ber}) is a special case of Equation~(\ref{eq:bound_gallager_fer}) for $k=1$. In the following, we will show by means of rate-distortion theory that the lower bounds in Equation~(\ref{eq:bound_gallager_ber}) and Equation~(\ref{eq:bound_gallager_fer}) can be met with equality by optimized channel coding schemes. We will denote the average frame error ratio measured in a system optimized w.r.t.\ to minimum $\BER$ by $\FER'$. Likewise, the average bit error ratio measured in a system optimized w.r.t.\ to minimum $\FER$ will be denoted by $\BER'$.

\section{Obtaining the lowest possible BER for a memoryless channel with given capacity}
\label{sec:lowest_ber}

We investigate the transmission of data at a rate which exceeds the
capacity, i.e.\ we consider the region where error-free transmission is
not possible. Rate-distortion theory \cite{shannon59a} postulates,
that if an end-to-end average bit error ratio $\BER_{\mathrm{T}}$
is tolerated, a code with rate $R$ and appropriate decoding rule
exists and achieves an average bit error ratio $\mathrm{BER} \le
\mathrm{BER}_\mathrm{T}$ as long as 

\begin{equation}
R \le \frac{C}{1-\mathrm{e}_2(\mathrm{BER}_\mathrm{T})}
\label{eq:rate_distortion_ber}
\end{equation}
and if $n\rightarrow \infty$.

We define a coding scheme (i.e. code, encoder, and decoder) with rate $R=\frac{C}{1-\mathrm{e}_2(\mathrm{BER}_\mathrm{T})}$ to be \emph{ideal in terms of the average bit error ratio}, iff the average bit error ratio meets the tolerated one, $\mathrm{BER} = \mathrm{BER}_\mathrm{T}={\mathrm{e}}_2^{-1}\left(1-\frac{C}{R}\right)$. It was shown in \cite{huettingeretal02} that Equation~(\ref{eq:rate_distortion_ber}) is met with equality when signaling over a memoryless binary symmetric channel (BSC).

This leads to the conclusion that the use of a coding scheme ideal w.r.t.\ $\BER$ results in an end-to-end channel which is a \emph{memoryless} BSC \cite{huettingeretal02}.
For completeness we add that in \cite{huettingeretal02} it is also shown that for an ideal coding scheme w.r.t.\ $\BER$, $\bar{I}(U;\hat{U}) \equiv \bar{I}(U;V)\equiv\frac{1}{k}I(\ve{U};\ve{V})$ holds, i.e.\ soft-output has no benefit
over hard-output and interleaving has no influence on the memoryless sequence of errors.

\section{Obtaining the lowest possible FER for a memoryless channel with given capacity}
\label{sec:lowest_fer}

In this section we present the first contribution of this paper. We discuss a channel coding system leading to the end-to-end channel with the lowest possible average frame error ratio. The adaptation of the converse to the channel coding theorem for the $\FER$ being the error ratio is given by Equation~(\ref{eq:bound_gallager_fer}).

\begin{figure}
\begin{center}
\psfrag{FER}[cb][cb]{$\FER$\, $\rightarrow$}
\psfrag{k}[ct][ct]{$k$\,$\rightarrow$}
\psfrag{All input Cap values Cap values Cap values}[l][l]{\small{$C/R=0.1$ ({\normalsize{$\circ$}}) to $C/R=0.9$ ($\lhd$)}}
\includegraphics[scale=0.45]{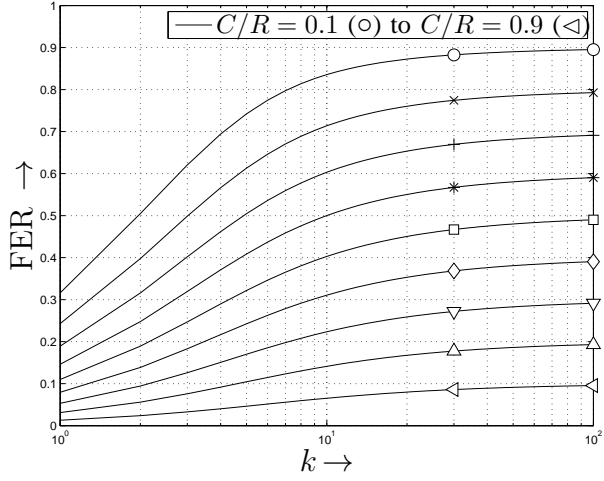}
\end{center}
\caption{\label{fig:FER_over_input_length_for_different_capacity}Minimum $\FER$ for different block lengths and given $C/R$}
\end{figure}

Figure \ref{fig:FER_over_input_length_for_different_capacity} shows the corresponding lower bound on the $\FER$ over $k$ for different values of $C/R$. 
The lower bound on $C/R$, specified by Equation~(\ref{eq:bound_gallager_fer}), becomes particularly interesting when $k$ approaches large values. Then, 
$C/R\geq\lim_{k\to\infty}\left(1-\frac{\efkt_{2}(\FER)}{k}-\frac{\ld(2^k-1)}{k}\FER\right)=1-\FER$,
%
or equivalently,

\begin{equation}
\FER\geq 1-C/R.
\label{eq:lower_bound_fer}
\end{equation}

Again, we show that this bound can be met with equality by means of rate-distortion theory and by this it is proven that a coding scheme meeting the $\FER$ has indeed to exist \cite{berger71}. 
To this end, let us first show that there exists a channel meeting inequality~(\ref{eq:lower_bound_fer}) with equality, namely the $M$-ary symmetric channel ($M$-SC) with $M=2^k$.
We assume that $\ve{u}$ and $\hat{\ve{u}}$ denote the input and output symbols of that channel, respectively. The transition probabilities of this end-to-end channel are denoted as $\Pr(\ve{U}=\ve{u}^{(j)}\mid\hat{\ve{U}}=\ve{u}^{(j)})=1-\FER$ $\forall j\in\{1,\dots,2^k\}$ and $\Pr(\ve{U}=\ve{u}^{(i)}\mid\hat{\ve{U}}=\ve{u}^{(j)})=\FER/(2^k-1)$, $\forall j$, $j\not=i$. 

%

The mutual information per channel use is calculated by
\begin{eqnarray}
\label{eq:capacity_mary_sc}
I(\ve{U};\hat{\ve{U}})\hspace{-0.2cm}&=&\hspace{-0.2cm}\left.H(\ve{U})-H(\ve{U}\mid \hat{\ve{U}})\right|_{\Pr(\ve{U})=[2^{-k}\dots 2^{-k}]}=\\
&\hspace{-2.3cm}&\hspace{-2.3cm}k+\sum\limits_{i=1}^{2^k}\sum\limits_{j=1}^{2^k}\Pr(\hat{\ve{U}}={\ve{u}}^{(j)}\mid \ve{U}=\ve{u}^{(i)})\Pr(\ve{U}=\ve{u}^{(i)})\ld\left(\frac{\Pr(\hat{\ve{U}}={\ve{u}}^{(j)}\mid \ve{U}=\ve{u}^{(i)})\Pr(\ve{U}=\ve{u}^{(i)})}{\sum\limits_{i'}\Pr(\hat{\ve{U}}={\ve{u}}^{(j)}\mid \ve{U}=\ve{u}^{(i')})\Pr(\ve{U}=\ve{u}^{(i')})}\right),\nonumber
\end{eqnarray}
what can be simplified to 

\begin{equation}
I(\ve{U};\hat{\ve{U}})=k-\efkt_{2^k}(\FER).
\label{eq:capacity_mary_sc_2}
\end{equation}
Equation~(\ref{eq:capacity_mary_sc_2}) denotes the mutual information for the transmission of a whole vector. Normalized to one binary symbol it reads 
$\frac{I(\ve{U};\hat{\ve{U}})}{k}=1+\frac{(1-\FER)}{k}\ld(1-\FER)+\frac{\FER}{k}\ld\left(\frac{\FER}{2^k-1}\right)$,
%
%
and thus for $k\to\infty$:

\[
\frac{I(\ve{U};\hat{\ve{U}})}{k}=1-\FER.
\]
Considering the data processing theorem in the form  $\frac{I(\ve{U};\hat{\ve{U}})}{k}\leq \frac{C}{R}$ and Equation (\ref{eq:lower_bound_fer}) in the form $C/R\geq 1-\FER$ allows us to conduct $\frac{I(\ve{U};\hat{\ve{U}})}{k}=\frac{C}{R}=1-\FER$ for $k\to\infty$. Again, by making use of rate-distortion theory and considering the fact that a distinct test chanel exists \cite{berger71}, we conclude that the lower bound provided in Equation~(\ref{eq:lower_bound_fer}) can be met with equality. 

In the following, we will review the properties of this channel.
In the case of an error, the $2^k$-ary symmetric channel maps the input to all incorrect outputs with equal probability. We therefore conclude that if a frame error occurs, the average bit error ratio within these frames is $0.5$. Hence, the resulting end-to-end channel corresponds to a fully bursty channel. More strictly speaking, a block-erasure channel with average erasure probability $\FER$ and infinite frame length meets the bound $C/R\geq 1-\FER$ with equality. This finding allows us to establish a coherence between the capacity and the rate, when an average frame error ratio $\FER_{\mathrm{T}}$ is tolerated. This coherence reads
$R\leq \frac{C}{1-\FER_{\mathrm{T}}}$.

We define a coding scheme (i.e. code, encoder, and decoder) with rate $R=\frac{C}{1-\mathrm{FER}_\mathrm{T}}$ to be \emph{ideal in terms of the average frame error ratio}, iff the average frame error ratio meets the tolerated frame error ratio $\mathrm{FER}_\mathrm{T}$ with equality, $\mathrm{FER} = \mathrm{FER}_\mathrm{T}$.

We denote the obtained average bit error ratio of such a channel by $\BER'$. There exists a straightforward coherence between $\BER'$ and the optimal average frame error ratio $\FER$ which reads $\BER'=\frac{1}{2}\left(1-\frac{C}{R}\right)$. The capacity of the fully bursty binary (end-to-end) channel, where all errors are part of very long error bursts, can be written as $C=1-2\BER'$. This is due to the fact that all errors are concentrated in bursts and within these bursts the average bit error ratio is $0.5$. Reliable communication is accomplished by the simple rule of erasing the error bursts at the receiver side. For error detection, e.g.\ by means of a cyclic redundancy check (CRC), additional redundancy is necessary but this cost vanishes for $k\to\infty$. Alike stated in Section \ref{sec:lowest_ber}, we observe that soft information has no benefit over hard output if an optimal coding scheme w.r.t.\ minimum $\FER$ is used. Examples for such fully bursty channels can simply be generated by renewal burst channel models, like the model of Fritchman with a single error state \cite{fritchman67}. For given $\BER$, the capacity of such a channel is maximized when the average burst length tends to infinity and in this limit, the capacity equals $1-2\BER$. This entity exactly corresponds to the situation of bit errors at the output of a coding scheme which is ideal w.r.t.\ minimum average frame error ratio.
Consider now an end-to-end channel with minimum $\BER$. For $k\to\infty$, the obtained average frame error ratio, denoted by $\FER'$ is given by $\FER'=0$ iff $\BER=0$ and $\FER'=1$ otherwise.
When considering blocks of infinite length, every block is erroneous if bit errors are possible in general.

With the results derived so far, it is straightforward to see that a channel coding scheme working in the region $R>C$ cannot obtain the minimum-possible $\BER$ and the minimum possible $\FER$ with one channel coding scheme, cf.\ Figure~\ref{fig:ber_and_ber_prime}.

%

\section{Possible Application}
\label{sec:possible_application}

A possible application of the presented results is introduced in this section. We assume binary antipodal signaling (BPSK) over the AWGN channel with a channel code of given rate $R$. A lower bound on the obtainable $\BER$ is given in Equation~(\ref{eq:bound_gallager_ber}), which can be rewritten to
$\BER\geq \efkt_{2}^{-1}\left(1-\frac{C}{R}\right)$.

\begin{figure}
\subfigure[Visualization of the significant difference between $\BER$ and $\BER'$ for coding schemes optimal w.r.t\ $\BER$ and $\FER$, respectively]{\label{fig:ber_and_ber_prime}
\psfrag{BER}[l][l][1]{\footnotesize{$\BER$}}
\psfrag{one minus c by r}[ct][ct]{\footnotesize{$1-C/R$ $\rightarrow$}}
\psfrag{BER p}[l][l][1]{\footnotesize{$\BER'$}}
\psfrag{BER and BER prime}[cb][cb][1]{\footnotesize{$\BER$ and $\BER'$ $\rightarrow$}}
\includegraphics[scale=0.48]{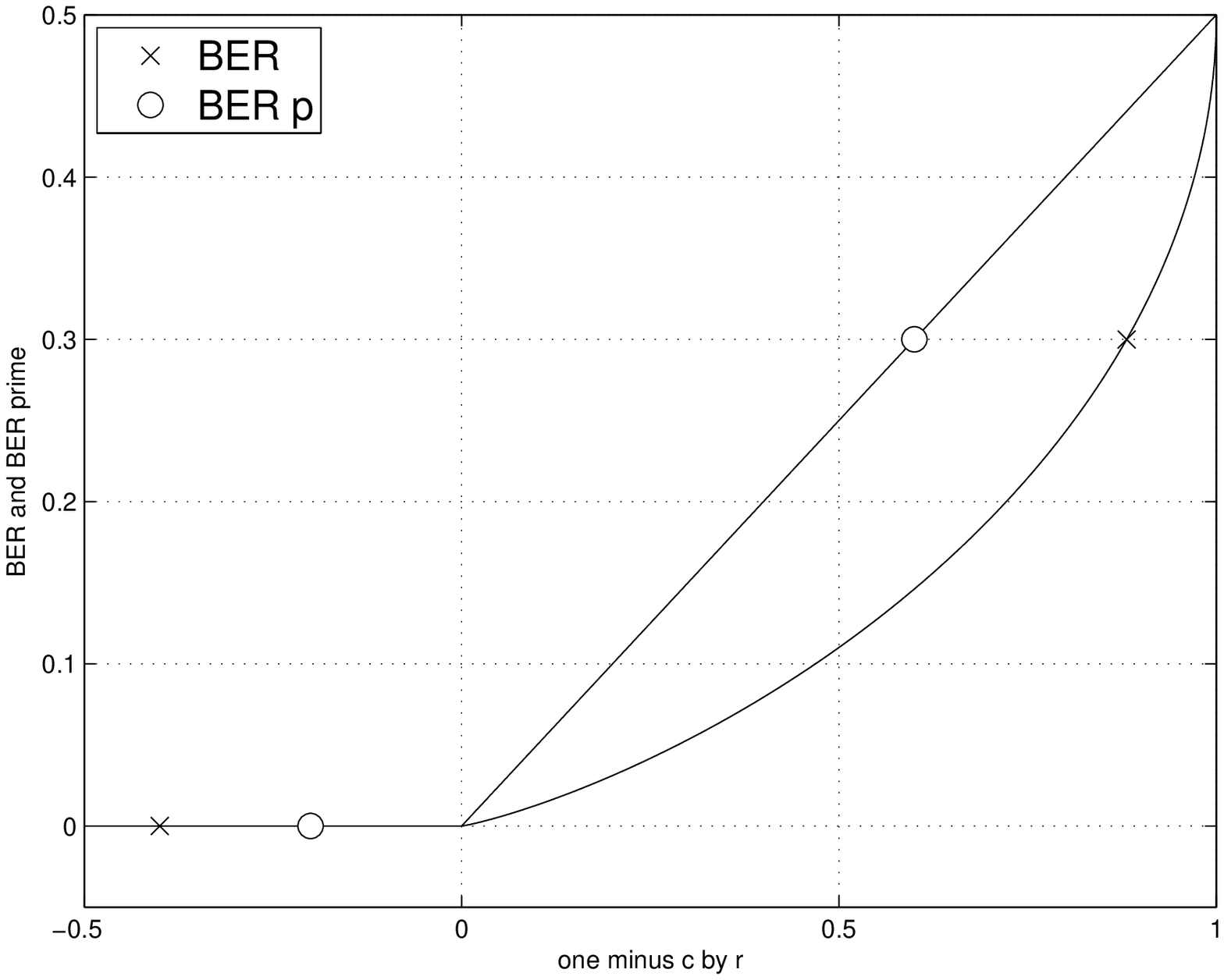}
}
\subfigure[$\BER$, $\BER'$, $\FER$ and $\FER'$ for coding schemes of given rate.]{\label{fig:ber_t_ber_prime_t_fer_t_fer_prime_t}
\psfrag{SNR}[ct][ct][1]{\footnotesize{$10\log_{10}(E_{\mathrm{b}}/N_0)$\,$\rightarrow$}}
\psfrag{xER}[cb][cb][1]{\footnotesize{$\BER$, $\BER'$, $\FER$\,$\rightarrow$}}
\psfrag{R1}[r][r]{\footnotesize{$R=\frac{1}{4}$}}
\psfrag{R2}[r][r]{\footnotesize{$R=\frac{1}{2}$}}
\psfrag{R3}[r][r]{\footnotesize{$R=\frac{3}{4}$}}
\psfrag{BER as well as BER prime_}[l][l][1]{\footnotesize{$\BER$ ($\times$) and $\BER'$ ($\circ$)}}
\psfrag{FER as well as FER prime_}[l][l][1]{\footnotesize{$\FER$ ($\times$) and $\FER'$ ($\circ$)}}
\includegraphics[scale=0.48]{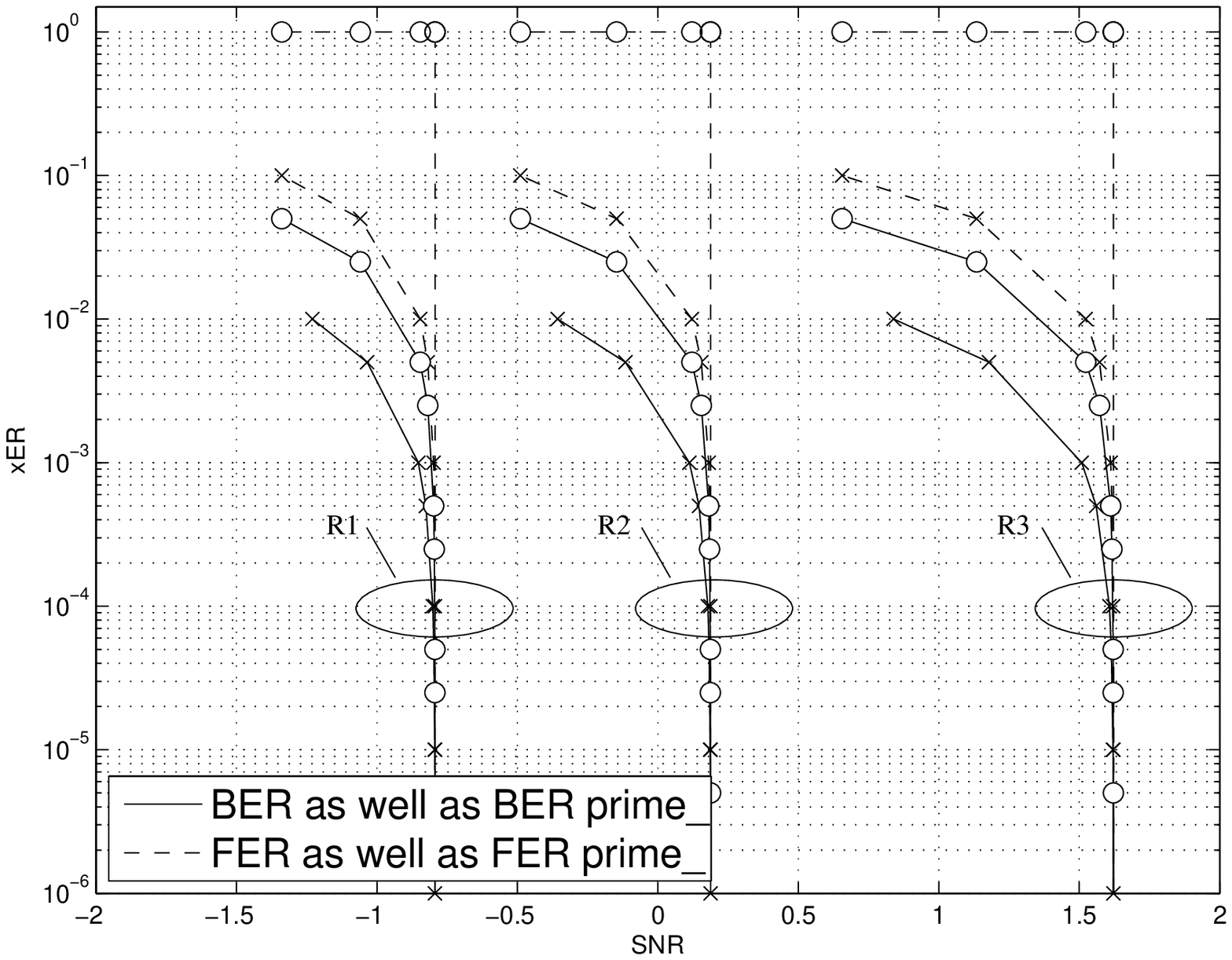}
}
\caption{Performance comparisons on $\BER$, $\BER'$, $\FER$ and $\FER'$}
\end{figure}

This entity allows to generate the curves depicting the optimum $\BER$ obtainable by codes of given rate and length approaching infinity. These curves are well-known from numerous publications within the area of channel coding and visualize the fundamental limits for transmission at a given rate. In Figure \ref{fig:ber_t_ber_prime_t_fer_t_fer_prime_t} these curves are shown for the rates $R=1/4$, $R=1/2$, $R=3/4$, respectively, and are labeled by $\BER$. Here, the capacity of the channel is specified by the signal-to-noise ratio expressed by $10\log_{10}(E_{\mathrm{b}}/N_0)$ as usual. In this context, $E_{\mathrm{b}}$ denotes the energy per transmitted bit of information and $N_0$ represents the one-sided spectral noise-power density.

The findings presented in Section~\ref{sec:lowest_fer} allow to extend
these fundamental curves to scenarios where the $\FER$ is used as the
performance measure. Here, Equation~(\ref{eq:lower_bound_fer}) states
the lower bound on the $\FER$ which can be reached by code lengths
approaching infinity. Figure~\ref{fig:ber_t_ber_prime_t_fer_t_fer_prime_t} also
depicts lower bounds on the frame error ratios for codes of
rate $R=1/4$, $R=1/2$, and $R=3/4$. These curves are labeled by $\FER$.
Figure~\ref{fig:ber_t_ber_prime_t_fer_t_fer_prime_t} also shows the
average error ratios $\BER'$ and $\FER'$. These curves illustrate the average bit error ratio for the case that the channel coding system of the end-to-end channel has been optimized for the $\FER$ and the average frame error ratio for the case that the optimization was done for the $\BER$, respectively. This confirms that
channel coding schemes being ideal w.r.t.\ minimum $\BER$ and minimum
$\FER$ have to be designed in quite different ways. Especially an optimization w.r.t.\ the $\BER$ in situations, when a low average frame error ratio is required as well, leads to significant performance losses. In all cases, the
average mutual information of the end-to-end channel is given by
$\min\left(C/R,1\right)$. In order to obtain minimum $\BER$, the
end-to-end channel is a memoryless BSC, whereas for minimum $\FER$, an
end-to-end channel with memory (to be precise, a block-erasure
channel) results.

\section{Conclusions}
\label{sec:conclusions}

We considered transmission at rates exceeding the capacity of
the underlying channel. Fundamental insight in a threefold manner
is given. First, knowledge on a coding scheme leading to an
end-to-end channel with minimum average frame error ratio is provided. It turns
out that this channel is a block-erasure channel, transmitting frames
either correctly or in such a way that no information is transmitted
at all. The average bit error ratio within a frame corresponding to a burst
error equals $0.5$.  Second, it was shown that minimum $\BER$ and
minimum $\FER$ are disparate requests to a channel coding scheme. The
third contribution is an application. It is usual in literature to
compare the $\BER$ behavior of channel coding schemes to information
theoretic bounds; this is now also possible with respect to the
$\FER$.

\section*{Acknowledgment}
{\emph{The authors want to thank the anonymous
    reviewers for their valuable comments.}}

\bibliography{LDPC_Group_Bibfile}

\end{document}